\documentclass[prl,twocolumn,showpacs]{revtex4}
\usepackage{graphicx}
\def\beq{\begin{equation}}
\def\eeq{\end{equation}}
\def \bea{\begin{eqnarray}}
\def\eea{\end{eqnarray}}

\def\d1av{{\overline d}_1}
\def\qav{{\overline q}}
\def\Df{D\!f}
\def\Dfresav{\overline{D\!f}^{\rm(res)}}
\def\Dfbg{D\!f^{\rm(bg)}}
\def\Thetasc{{\hat\Theta}}
\def\psireg{\psi_{E}^{\rm{(reg)}}}
\def\psiirr{\psi_{E}^{\rm{(irr)}}}

\begin{document}

\title{Statistical Properties of Fano Resonances in Atomic and 
Molecular Photoabsorption}

\author{Wolfgang Ihra
       \medskip}

\affiliation{Theoretical Quantum Dynamics, 
             Universit\"at Freiburg, 
             Hermann-Herder-Str.~3, 
             D-79104 Freiburg, Germany}

\begin{abstract}
  Statistical properties of Fano resonances occurring in 
  photoabsorption to highly excited atomic or
  molecular states are derived. The situation with one open 
  and one closed channel is analyzed when
  the classical motion of the excited complex in the closed
  channel is chaotic. The closed channel subspace is modeled by
  random matrix theory. The probability distribution of the Fano
  parameter is derived both
  for the case of time reversal symmetry (TRS) and broken time 
  reversal symmetry. For the TRS case 
  the area distribution under a resonance profile relevant for
  low resolution experiments is discussed in detail.
\end{abstract}
\maketitle
{\it Introduction\/} ---
Resonance phenomena are ubiquitous in highly excited quantum systems.
Typical examples are auto-ionizing atomic resonances
\cite{Fri98} or photodissociation of rovibrational molecular states 
\cite{Schi93}. When the density of resonances is high a
common and natural way to extract information about 
resonance properties is to resort to statistical
methods \cite{Haake01,Stoeck99,Guhr98}. 
Typically as a consequence of the high excitation the resonant complex
behaves chaotically from a classical point of view. 
It is then appropriate to model the excited complex by
means of random matrix theory \cite{BGS84,Alhas92,Alhas98}. 

Here
I
address a problem frequently encountered in photoabsorption
processes of atomic and molecular systems: What are
the statistical properties of line shapes when photoabsorption
into a highly excited resonant state takes place?
Since the seminal work of Fano \cite{Fano61} it is known that when
at least two pathways exist for an atomic or molecular system to
decay after having absorbed a photon the line shapes of resonances
differ in general considerably from the Breit-Wigner form. In
photodissociation of molecules or photoexcitation of an atom
to an auto-ionizing resonance interference between 
the indirect and the direct decay process results in 
a Beutler-Fano resonance profile of the photoabsorption cross section 
\cite{Fri98,Conn88,Durand01} (abbreviated as Fano profile in the following). 
The shape of the resonant part of the cross section
is determined by a single parameter, the Fano parameter $q$.
This contribution is aimed at such a statistical theory of Fano resonances.

It is worthwhile to stress that atomic and molecular systems 
are ideally suited to test the predictions on Fano resonances
as neither Coulomb or dephasing effects which may become
relevant in mesoscopic structures have to be taken into account \cite{Clerk01}.
\smallskip

\begin{figure}[b]
  \centerline{\includegraphics[width=7.0cm]{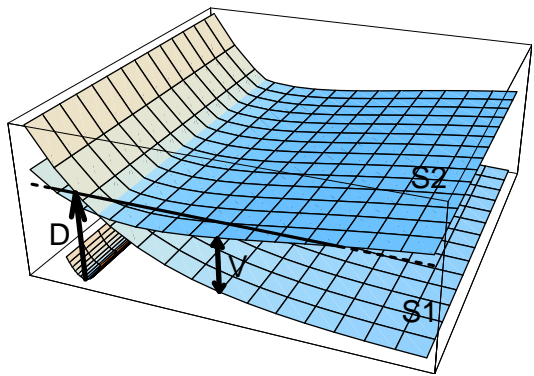}}
      \caption[Short title]{\label{fig:fig1}
         Schematic sketch of the photodissociation
         process: Laser excitation takes place from a low lying electronic 
         potential surface (dipole operator $D$).
         The classical dynamics on surface S2 is bound and chaotic.
         Dissociation can take place directly by dipole
         excitation to channel 1 or indirectly by excitation to channel
         2 and the coupling $V$ between the surfaces S1 and S2.
}
\end{figure}

{\it Theory\/} --- For photodissociation of a molecule 
Fig. \ref{fig:fig1} schematically depicts the situation envisaged
\cite{footnt1}. 
The molecule is coherently laser excited from the ground state or a 
low lying state $|0\rangle$ of energy $E_0$
onto two electronic potential surfaces.
The transition is accomplished by the dipole operator $D$.
Electronic surface S1 is the open channel --- the energy $E$ of the 
molecular complex is above the dissociation threshold --- while the motion
on the potential surface S2 is bound (closed channel). The dipole
matrix element for transition to the open channel is given by 
$d_1\equiv\langle 0|D|\psireg\rangle$, where $|\psireg\rangle$ is the 
regular continuum solution at energy $E$ in the open channel,
normalized in energy. When $E$ coincides with the energy $E_n$ 
of a bound state $|\phi_n\rangle$ in the channel 2 the
closed channel carries the transition amplitude 
$\langle 0|D|\phi_n\rangle$ in the absence of coupling to the
continuum. In the following it is 
assumed that transitions between the two excited manifolds are possible.
In the diabatic representation the coupling
between the two surfaces is given by a non-diagonal potential $V$. 
Then the eigenstate $|\phi_n\rangle$ turns into a resonance with width
$\Gamma_n=2\pi|\langle\psireg|V|\phi_n\rangle|^2$.

In the regime $\Gamma_n\ll\Delta$ of isolated resonances ($\Delta$
being the mean energy spacing of resonances) the oscillator strength $\Df(E)$
for the dipole transition is given by \cite{Fri98}
\beq\label{eq:oscstrength}
  \Df(E) = \Dfbg \, \frac{|q_n+\epsilon_n|^2}{1+\epsilon_n^2} \,,
\eeq
where $\Dfbg \equiv 2\mu (E-E_0)|d_1|^2/\hbar^2$ is the 
background oscillator strength for the transition to the open channel
in the absence of coupling to the closed channel, $\mu$ is the reduced
mass of the excited complex 
and $\epsilon_n\equiv2(E-E_n-\Delta_n)/\Gamma_n$ the reduced energy.
The energy shift $\Delta_n$ of the
resonance is neglected in the following. The line shape of the resonance
is parameterized by the Fano parameter $q_n$. For a 
non-degenerate continuum it is given by \cite{Fri98}
\beq\label{eq:qpar}
  q_n = \frac{\langle 0|D|\phi_n\rangle}
        {\pi d_1\langle\psireg|V|\phi_n\rangle} + \qav
     \,,\quad
  \qav = -\frac{\langle 0|D|\psiirr\rangle}
    {\langle 0|D|\psireg\rangle} \,.
\eeq
The irregular continuum solution $|\psiirr\rangle$ appears in the
expression for $q_n$ because near resonance the wave function in the open channel
is a superposition of the regular and the irregular solution.
Both $d_1$ and $\qav$ depend 
only weakly on $E$ and are assumed to be constant within the  
energy window over which a sample of resonances is taken. 
From (\ref{eq:qpar}) it is seen that 
the distribution of a set of Fano parameters
$\{q_n\}$ taken over a stretch of the energy spectrum 
is determined by the statistical properties of the set of the
eigenstates $\{|\phi_n\rangle\}$ of the closed channel. 

The distribution of the Fano parameter is derived under the two 
following conditions: Firstly the classical motion of the excited
molecular complex on the electronic surface S1 of the closed channel
is completely chaotic. This ensures that the statistical properties
of generic wave functions in chaotic systems apply and that the closed
channel subspace can be modeled by random matrix theory \cite{Haake01}.
Secondly the excitation 
process and the coupling between the two electronic surfaces are
assumed to be spatially well separated: the overlap of the initial wave packet 
$\langle {\bf r}|D|0\rangle$ and the coupling potential $V({\bf r})$ 
in coordinate representation is negligible. Therefore
$x\equiv\langle 0|D|\phi_n\rangle/d_1$ and
$y\equiv\pi\langle\psireg|V|\phi_n\rangle$
can be taken statistically independent random variables \cite{footnt2}.
\medskip

{\it Distribution of the Fano parameter\/} ---
With these ingredients the calculation of the probability distribution
$P(q)$ of the Fano parameter is straightforward. (The index $n$ of the
resonance is omitted in the following.)
In the case of time reversal symmetry
the wave functions and viz. $x$ and $y$ can be chosen real.
The closed channel subspace is modeled by $N\times N$ matrices
taken from the Gaussian orthogonal ensemble (GOE). 
All results are understood in the limit $N\to\infty$ where $x$ and $y$ are
Gaussian random random variables with zero mean and 
variances $\sigma_x^2$ and $\sigma_y^2$.
The probability distribution is given by 
\beq\label{eq:PGOE}
  P_{\rm GOE}(q) = \!\int\limits_{-\infty}^{\infty}\! dx \!
                   \int\limits_{-\infty}^{\infty}\! dy \,
                P_{\sigma_x}(x) \,
                P_{\sigma_y}(y) \,
                \delta\!\left(q-\qav-\frac{x}{y}\right) 
\eeq
and $P_{\sigma_x}(x)$ is the probability distribution of $x$ (and likewise
for $y$). The integral is most easily evaluated by using the Fourier 
integral representation of the delta function and the result is
\beq\label{eq:PGOEfin}
   P_{\rm GOE}(q;s) = \frac{1}{\pi}
             \frac{s}{s^2+(q-\qav)^2} \,,\quad\quad
    s\equiv\sigma_x/\sigma_y \,.     
\eeq
The probability distribution of the Fano parameter in the GOE case thus
turns out to be Lorentzian with mean value $\qav$
and width $s$. 

The width $s$ of the probability distribution is related to the
coupling strength $V$ between the closed channel and the continuum 
channel and the ratio of the dipole transition matrix elements to
both channels. Assume $V$ can be written in the form $V=\lambda V_0$
where $\lambda$ characterizes the coupling strength and $V_0$ is
fixed. Then $\sigma_y\sim\lambda$ and therefore $s\sim1/\lambda$.
For strong coupling to the continuum the Lorentz distribution acquires
a small width centered around $\qav$. The same holds
if direct photoexcitation dominates over the indirect process
since then $\sigma_x$ becomes small.

In the case of broken time reversal symmetry the Hamiltonian of the 
closed channel is modeled by $N\times N$ matrices from the Gaussian
unitary ensemble (GUE). In this case $x$ and $y$ are complex Gaussian 
random variables with independent real and imaginary parts and 
the Fano parameter is in general complex. Most conveniently the 
distribution of $q$ is characterized
by the probability distribution of the phase 
$\varphi_q$ and the modulus $r_q$ of the quantity $q-\langle q\rangle$. The phases
$\varphi_x$ of $x$ and $\varphi_y$ of $y$ are uniformly distributed 
${\rm mod}(2\pi)$ and the same holds for $\varphi_q=\varphi_x-\varphi_y$.
Denoting the modulus of $x$ by $r_x$ and of $y$ by $r_y$ 
the probability distribution $P_{\rm GUE}(r_q)$ is given by
\beq\label{eq:PGUE}
   P_{\rm GUE}(r_q) = \! \int\limits_{0}^{\infty} dr_x \!
                      \int\limits_{0}^{\infty} dr_y \,
                     P(r_x)\,  P(r_y) \,
               \delta\!\left(r_q-\frac{r_x}{r_y}\right) \,,
\eeq
where $r_x$ (and likewise $r_y$) has the probability distribution
$P(r_x)=\sigma_{r_x}^{-2} r_x \exp(-r_x^2/2\sigma_{r_x}^2)$. [Notice
that $P(r_x^2) \sim \exp(-r_x^2/2\sigma_{r_x}^2)$ has the form
of a Porter-Thomas distribution for GUE.]  
Proceeding as in the GOE case the integrals turn out to be more
complicated as they involve error functions. The final result is 
nevertheless simple and reads
\beq\label{eq:PGUEfin}
   P_{\rm GUE}(r_q) = \delta(r_q) + 
     \frac{s_r^2 r_q}{\left(s_r^2+r_q^2\right)^2} \,,
     \quad\quad
     (r_q\ge 0) \,,
\eeq
where $s_r\equiv\sigma_{r_x}/\sigma_{r_y}$. The distribution
is delta peaked at $r_q=0$ and has a local maximum at
$r_q^{(0)} = s_r/\sqrt{3}$ with $P(r_q^{(0)})= 3\sqrt{3}/(16s_r)$.
Again, as discussed for the GOE case the width $s_r$ of the probability
distribution $P_{\rm GUE}(r_q)$ 
is determined by the strength of the coupling 
between the two channels and the ratio of the strength
between the dipole transition to the closed and to the open channel.
\smallskip

\begin{figure}
  \includegraphics[width=7.0cm]{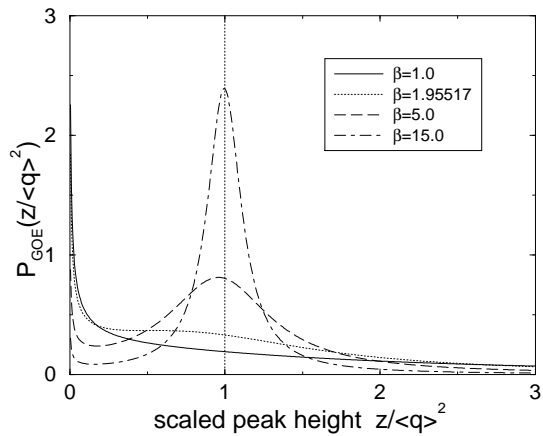}
  \caption[Short title]{\label{fig:fig2}
             Peak height distribution $P_{\rm GOE}(z/\qav^2)$
             in the GOE case
             for different values of $\beta=\qav/s$,
             $s$ being the width of the $q$-parameter distribution Eq.~(4).
             For $\beta\ge 1.95517$ the distribution
             displays a local maximum between zero and one.
}
\end{figure}

{\it Peak height distribution for GOE\/} ---
For the TRS case the maximum of the Fano profile is given at
$\epsilon=1/q$ with peak height $z\equiv[\Df-\Dfbg]/\Dfbg=q^2$ 
relative to the background oscillator strength. The distribution
of peak heights can easily be calculated from 
(\ref{eq:PGOEfin}) and is given by
\beq\label{eq:peakdist}
   P_{\rm GOE}(z) = \frac{1}{2\sqrt{z}}\left[
      P_{\rm GOE}(\sqrt{z};s)
    + P_{\rm GOE}(-\sqrt{z};s)
   \right]
\eeq
with $z>0$. In Fig. \ref{fig:fig2} 
$P_{\rm GOE}(z)$ is plotted in terms of the scaled 
variable $z/{\qav}^2$ and different values of the parameter
$\beta=\qav/s$. It has a local maximum at a value $0<z/\qav^2<1$ 
for $\beta\ge 1.95517$. Therefore for a local maximum to occur the
width $s$ of the Lorentz profile (\ref{eq:PGOEfin})
must be smaller than approximately twice the average 
value $\qav$ of the Fano parameter. For $\beta\to\infty$ 
the position of the maximum approaches $z=\qav^2$.
\medskip

{\it Profile area distribution for GOE\/} --- 
In low resolution experiments the relevant experimental quantity 
is the profile area distribution under the resonance rather
than the peak height distribution. To be more specific assume
that the band width $\delta E$ of the exciting laser beam is larger 
than the
width $\Gamma$ of the resonance but still smaller than the mean 
spacing $\Delta$ between adjacent resonances 
($\Gamma < \delta E < \Delta$). Assuming a rectangular laser
profile the excess oscillator strength
averaged over the resonance profile
with respect to the background strength is given by \cite{Bohm86} 
\beq\label{eq:areadef}
     \Dfresav(E) = \frac{1}{\delta E} \!\!
            \int\limits_{E-\delta E/2}^{E+\delta E/2} \!\!
              \Dfbg(E') \left[F(q;\epsilon)-1 \right] dE'        
\eeq
where
$F(q;\epsilon)\equiv|q+\epsilon|^2/(1+\epsilon^2)$
is the profile function of the resonance [cf. (\ref{eq:oscstrength})]. 
Since $\Gamma < \delta E$ the range
of integration in (\ref{eq:areadef}) can be extended to infinity and 
$\Dfresav(E)$ is given by
\beq\label{eq:area2}
   \Dfresav(E) = \Dfbg \frac{\Gamma}{2\,\delta E}
                   \int\limits_{-\infty}^{\infty}
                   \frac{q^2+2 q\epsilon-1}{1+\epsilon^2}
                    d\epsilon \,,
\eeq 
which performing the integration gives
$\Dfresav/\Dfbg=\pi\Gamma (q^2-1)/2 \equiv \Theta$.
In the following the statistical distribution of the observable
$\Theta$ is discussed. It can be written as $\Theta = x^2
+2\qav xy + (\qav^2-1)y^2\equiv f(x,y)$ in terms of the matrix elements $x$,
$y$ and the mean value $\qav$ of the Fano parameter. The probability
distribution of $\Theta$ is given by
\beq\label{eq:areaprob}
  P(\Theta) = \!\! \int\limits_{-\infty}^{\infty}\! dx \,
                 P_{\sigma_x}(x) \!
                \int\limits_{-\infty}^{\infty} \! dy \,
                P_{\sigma_y}(y) \,
                \delta\!\left(\Theta\!-\!f(x,y)\right) \,.
\eeq
Introducing the scaled area $\Thetasc\equiv a\Theta$ with 
$a = (\sigma_x\sigma_y)^{-1}\sqrt{1/4+b^2}$
and
$b = \left[s+s^{-1}(\qav^2-1)\right]/4$
the distribution $P(\Thetasc)$ can be written as
\beq\label{eq:areadistscal}
  P(\Thetasc) = \frac{1}{\pi\sqrt{1+4 b^2}} \exp(\mu\Thetasc)
     K_0(|\Thetasc|) \,,
\eeq
where $K_0$ is the MacDonald function.
Additionally the {\it asymmetry parameter\/}
$\mu = b\,[\frac{1}{4}+b^2]^{-1/2}$ has been introduced
which determines how much $P(\Thetasc)$ deviates from a symmetric 
distribution. $P(\Thetasc)$ diverges logarithmically as $\Thetasc$
approaches zero.

\begin{figure}
  \includegraphics[width=7.0cm]{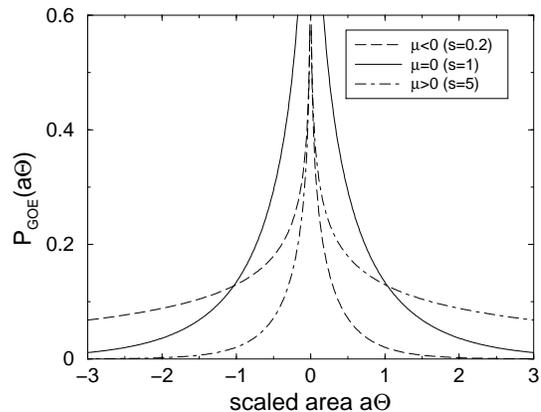}
  \caption[Short title]{\label{fig:fig3}
                        Distribution $P({\Thetasc})$ of the
                        scaled resonance area $\Thetasc=a\Theta$,
                        see discussion after Eq.~(10).
                        The average value
                        of the Fano parameter is fixed at $\qav = 0$ and 
                        $s$ is varied.
}
\end{figure}

Fig. \ref{fig:fig3} demonstrates the dependence of $P(\Thetasc)$
at fixed $\qav=0$ in the three characteristic regimes $\mu<0$,
$\mu=0$ and $\mu>0$ by varying $s$. For $\mu=0$ the probability distribution is 
symmetric with respect to $\Thetasc=0$ (solid line). For $\mu<0$ the
area distribution is asymmetric with the flat side at negative values of 
$\Thetasc$ (dashed line). The opposite holds for $\mu>0$  
(dot-dashed line). 

\begin{figure}
   \centerline{\includegraphics[width=7.5cm]{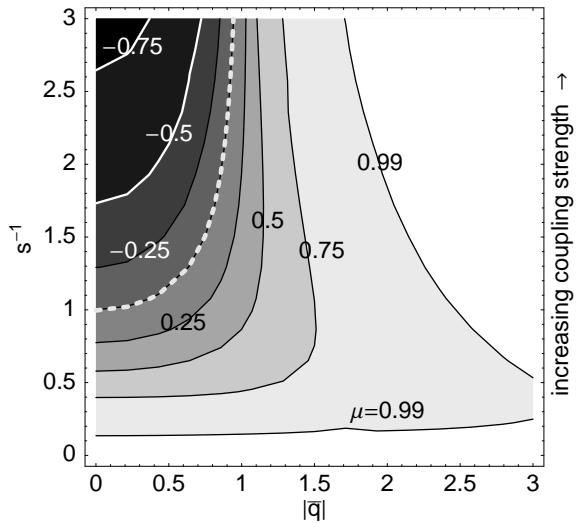}}
  \caption[Short title]{\label{fig:fig4}
                       Asymmetry parameter $\mu$ of the area distribution,
                       Eq.~(11) as a function
                       of the average shape parameter $\qav$
                       and $s^{-1}$.
                       The white dashed line marks the values
                       $(\qav,1/s)$ where $\mu=0$.
}
\end{figure}

As $\mu$ depends both on $s$ and $\qav$ a contour plot of $\mu$
as a function of $(\qav,1/s)$ is presented in Fig.~\ref{fig:fig4}.
The variable $1/s$ instead of $s$ has been chosen for matters of
discussion as
$1/s\to\infty$ corresponds to the limit of strong coupling between
the closed and the continuum channel. The range of $\mu$ is $-1<\mu<1$. 
The white dashed line marks the subspace of parameters where
$P(\Thetasc)$ is symmetric $(\mu=0)$, given by $\qav^2 + s^2 = 1$.
If $|\qav|\ge 1$ the asymmetry parameter $\mu$ is always positive regardless
of $1/s$. For fixed $|\qav|<1$ and weak coupling ($1/s$ small)
the asymmetry parameter $\mu$ is positive, too. If the coupling
is enlarged $\mu$ becomes negative above a certain value of $1/s$ 
which depends on $\qav$. 

The expectation value of the scaled area distribution is given by
$\langle\Thetasc\rangle = \mu\,(1+4 b^2)$. Thus
$\langle\Thetasc\rangle$ is always positive when $|\qav|\ge 1$. 
Note that the area under an {\it
individual\/} resonance is positive if for its Fano parameter $|q|>1$ holds.
When $|q|<1$ the area associated with an individual resonance is negative. In
contrast $\langle\Thetasc\rangle$ 
can either be positive or negative for $|\qav|<1$ depending
on the coupling strength to the continuum.
Strong coupling to the continuum ($1/s\to\infty$) results in a
negative value of $\langle\Thetasc\rangle$. In the limit
$\mu\to+1$ Breit-Wigner resonances
dominate the spectrum. For fixed $|\qav|>1$ this limit is reached 
both for weak and large coupling to the continuum.
For $|\qav|<1$ the Breit-Wigner limit is reached 
only for weak coupling ($1/s\to 0$). For $|\qav|<1$ and 
strong coupling to the continuum ($1/s\to\infty$) the
expectation value of the area distribution is negative and 
window resonances with negative area dominate the spectrum
in the limit $\mu\to-1$.
\medskip

{\it Summary} --- Statistical properties of Fano
resonances were derived for the situation of competing direct
and indirect photoabsorption on the basis that the classical motion in the
closed channel is chaotic. This situation is frequently
encountered in photodissociation of highly excited 
rovibrational states of molecules 
or photoabsorption of atoms to auto-ionizing
resonances. The closed channel subspace is modeled by
random matrix theory. The distribution of the Fano parameter $q$
was derived for both the GOE and the GUE cases. The peak height
distribution and the resonance 
area distribution were discussed in detail for the GOE. 
\smallskip

\begin{acknowledgments}
{\it Acknowledgments\/} ---
I am grateful to John S. Briggs for pointing out to
me the problem of Fano resonances in classically
chaotic systems.
I am indebted to Thomas Seligman for discussions 
and his hospitality during a visit at CIC, Cuernavaca, Mexico,
where part of this work was completed.
Discussions with J. Flores, T. Gorin, B. Mehlig and M. M\"uller
are appreciated.
Financial support from SFB 276 (``Correlated dynamics
of highly excited atomic and molecular systems'') is acknowledged.
\end{acknowledgments}

%

\begin{thebibliography}{00}
\bibitem{footnt1} The situation is analogous for photoexcitation
                  of an auto-ionizing atomic resonance when the
                  system can be described by a model with
                  one open channel and one closed chaotic channel. 
                  An example is the hydrogen
                  atom in a homogeneous magnetic field in the chaotic regime 
                  between the first and the second Landau threshold
                  \cite{Gremaud93}.
\bibitem{footnt2} The situation where $x$ and $y$ are correlated is
                  deferred to a future publication.
\end{thebibliography}
\end{document}